\documentclass[prb,preprint,superscriptaddress]{revtex4-1}

\usepackage{amsmath,braket,SIunits,graphicx}

\newcommand{\ee}{\ensuremath{\textrm{e}}}
\newcommand{\ii}{\ensuremath{\mathrm{i}}}
\newcommand{\dd}{\ensuremath{\textrm{d}}}
\renewcommand{\vec}[1]{\ensuremath{\mathbf{#1}}}
\newcommand{\op}[1]{\ensuremath{\hat{#1}}}

\begin{document}

\title{Transition probability functions for inelastic electron--electron scattering}
\author{Stefan L{\"o}ffler}
\email{stefan.loeffler@tuwien.ac.at}
\affiliation{Institute of Solid State Physics, Vienna University of Technology, Wiedner Hauptstra{\ss}e 8-10, 1040 Wien, Austria}
\author{Peter Schattschneider}
\affiliation{Institute of Solid State Physics, Vienna University of Technology, Wiedner Hauptstra{\ss}e 8-10, 1040 Wien, Austria}
\affiliation{University Service Center for Transmission Electron Microscopy, Vienna University of Technology, Wiedner Hauptstra{\ss}e 8-10, 1040 Wien, Austria}

\begin{abstract}
In this work, the transition matrix elements for inelastic electron--electron scattering are investigated. The angular part is given by spherical harmonics. For the weighted radial wave function overlap, analytic expressions are derived in the Slater-type and the hydrogen-like orbital models. These expressions are shown to be composed of a finite sum of polynomials and elementary trigonometric functions. Hence, they are easy to use, require little computation time, and are significantly more accurate than commonly used approximations.
\end{abstract}

\maketitle

\section{Introduction}

Ever since \citeauthor{PRSLA_v82_i557_p495} shot $\alpha$ particles on a gold foil \cite{PRSLA_v82_i557_p495} and \citeauthor{PMS6_v21_i125_p669} subsequently used their results to confirm that atoms consist of a very small nucleus surrounded by an electron cloud \cite{PMS6_v21_i125_p669}, scientists have used scattering effects to determine the properties of otherwise invisible or inaccessible objects.

In particular, quantum systems are usually investigated by means of scattering experiments. The central quantity in these quantum mechanical scattering systems is the matrix element of the transition operator $\op{V}$,
\begin{equation}
	\bra{f} \braket{F | \op{V} | I} \ket{i},
\end{equation}
which describes the amplitude for the transition from a product state $\ket{I}\ket{i}$ to another product state $\ket{F}\ket{f}$ in first order Born approximation. Here, we distinguish between the states of the probe (lower-case letters) and the target (upper-case letters).\footnote{Of course, a scattering event can also produce entangled states. Since these can be decomposed into a linear combination of product states, the formalism presented here can be applied to those states as well.}

In this work, we limit ourselves to the treatment of the electronic transition of a single, isolated atom, and the effect it has on the probe triggering that transition. Assuming a Coulomb-like transition potential,
\begin{equation}
\begin{aligned}
	\bra{\vec{r}'} \braket{\vec{R}' | \op{V} | \vec{R}} \ket{\vec{r}} 
	&\propto \frac{1}{|\vec{r}-\vec{R}|} \delta(\vec{r}'-\vec{r}) \delta(\vec{R}'-\vec{R}) \\
	\bra{\vec{k}'} \braket{\vec{R}' | \op{V} | \vec{R}} \ket{\vec{k}} 
	&\propto \frac{\ee^{\ii (\vec{k}-\vec{k}')\cdot\vec{R}}}{|\vec{k}-\vec{k}'|^2} \delta(\vec{R}'-\vec{R}) \\
	&= \frac{\ee^{\ii \vec{Q}\cdot\vec{R}}}{Q^2} \delta(\vec{R}'-\vec{R})
\end{aligned}
\end{equation}
with $\vec{Q} := \vec{k}-\vec{k}'$ and inserting appropriate identity operators, the transition matrix element can be written as
\begin{equation}
\begin{aligned}
	& \bra{f} \braket{F | \op{V} | I} \ket{i} \\
	=& \int_{\vec{R},\vec{R}',\vec{k}.\vec{k}'} \braket{f | \vec{k}'} \braket{F | \vec{R}'} \bra{\vec{k}'} \braket{\vec{R}' | \op{V} | \vec{R}} \ket{\vec{k}} \braket{\vec{R} | I}\braket{\vec{k} | i} \\
	=& \int_{\vec{k},\vec{k}',\vec{R}} \frac{1}{Q^2} \braket{f | \vec{k}'} \braket{F | \vec{R}} \ee^{\ii \vec{Q}\cdot\vec{R}} \braket{\vec{R} | I}\braket{\vec{k} | i}.
\end{aligned}
\end{equation}
Here, we assume that the probe electron was prepared as a plane wave, i.e., $\ket{i} = \ket{\vec{k}}$, and that the detector also measures plane waves, i.e., $\ket{f} = \ket{\vec{k}'}$.\footnote{$\ket{i}$ and $\ket{f}$ can always be described in a plane wave basis. In the most general case, each is given by a coherent superposition of several plane wave components. Thus, the formulas derived in this article apply to each component of the general case as well.} In that case, the transition matrix element assumes its commonly used form,
\begin{equation}
	\frac{1}{Q^2} \int_{\vec{R}}  \braket{F | \vec{R}} \ee^{\ii \vec{Q}\cdot\vec{R}} \braket{\vec{R} | I},
\end{equation}
or, by virtue of the partial wave or Rayleigh expansion,
\begin{equation}
	\frac{4\pi}{Q^2} \sum_{\lambda,\mu} \ii^\lambda Y_\lambda^\mu(\vec{Q}/Q)^* \int_{\vec{R}}  \braket{F | \vec{R}} Y_\lambda^\mu(\vec{R}/R) j_\lambda(QR) \braket{\vec{R} | I},
	\label{eq:matrix_rayleigh}
\end{equation}
where the $Y_\lambda^\mu$ denote the spherical harmonics and the $j_\lambda$ are the spherical Bessel functions of first kind.

Up to this point, the treatment is exact. In order to evaluate the matrix element further, one needs to explicitly specify the functions $\braket{\vec{R} | I}$ and $\braket{\vec{R} | F}$, however. For the electronic transitions treated here, two common approximations are Slater-type orbitals (STO) \cite{PR_v36_i1_p57,TJoCP_v38_i11_p2686} and hydrogen-like orbitals (HLO) \cite{Egerton1996}. In both models, the initial and final states are modeled as a spherical harmonic angular dependence and a (analytic) radial wavefunction $\psi$:
\begin{equation}
	\braket{\vec{R} | I} = \psi_I(R) Y_{l}^{m}(\vec{R}/R) \qquad
	\braket{\vec{R} | F} = \psi_F(R) Y_{l'}^{m'}(\vec{R}/R).
\end{equation}
With this, eq.~\ref{eq:matrix_rayleigh} becomes
\begin{equation}
\begin{aligned}
	\frac{4\pi}{Q^2} \sum_{\lambda,\mu} \ii^\lambda Y_\lambda^\mu(\vec{Q}/Q)^* 
	&\int_{\Omega} Y_{l'}^{m'}(\Omega)^* Y_\lambda^\mu(\Omega) Y_{l}^{m}(\Omega) \dd^2\Omega \\
	\cdot&\int_0^\infty \psi_F(R) j_\lambda(QR) \psi_I(R) R^2 \dd R,
\end{aligned}
\end{equation}
where the angular integral evaluates to Wigner 3j symbols\cite{Nelhiebel1999,PRB_v59_i20_p12807,N_v441_i_p486}. Hence, we will only deal with the radial integral
\begin{equation}
	\langle j_\lambda(Q) \rangle := \int_0^\infty \psi_F(R) j_\lambda(Q R) \psi_I(R) R^2 \dd R
	\label{eq:radial_wf_overlap}
\end{equation}
in this work, for which we will derive an analytical form in the STO and HLO models.

The rationale behind this choice of orbitals is the fact that the initial state is usually a tightly bound (and hence strongly localized) core state that is described well by an atomic model. Since the matrix element can only be non-zero if both the initial and final states are non-zero, this effectively selects a portion of the final state that is close to the nucleus as well.

Crystal and many-body effects, on the other hand, are often relatively small perturbations---especially in the commonly experimentally accessible regions. In addition, for elucidating the underlying fundamentals, an analytical treatment in an isolated atom model is usually advantageous.

In many articles, eq.~\ref{eq:radial_wf_overlap} is simplified further by taking the small angle limit (also known as dipole-approximation),
\begin{equation}
	j_\lambda(Q R) \approx \frac{(QR)^\lambda}{(2\lambda+1)!!},
	\label{eq:jlambda_smallq}
\end{equation}
in which case the integral over $R$ in eq.~\ref{eq:radial_wf_overlap} becomes only a weighting factor (the factor $Q^\lambda$ can be moved out of the integral).

For increasing momentum transfer, a leading term $Q^\lambda$ would increase boundlessly, giving rise to a similar effect as the ultraviolet catastrophe. To avoid that, an artificial cutoff $Q_c$ is sometimes introduced, which is only fighting the symptoms instead of the cause, however, and is not very elegant. In addition, is has recently been shown experimentally using electron energy loss spectrometry (EELS) that a $Q^\lambda$ dependence is an oversimplification even for $Q < Q_c$ and can lead to errors of the order of \unit{25}{\%} \cite{U_v111_i_p1163,U_v80_i3_p183}.

\section{Slater-type orbitals}

The radial part of STOs is given by\cite{PR_v36_i1_p57,TJoCP_v38_i11_p2686}
\begin{equation}
	N R^{n - 1} \ee^{-\frac{\zeta R}{a_\mu}},
\end{equation}
where $n$ is an effective quantum number (which is not necessarily an integer). $\zeta := (Z - s)/n$ is an effective nuclear charge, with the physical nuclear charge $Z$ and a screening factor $s$. The normalization constant is given by
\begin{equation}
	N = \sqrt{\frac{(2\zeta)^{2n + 1}}{a_\mu \Gamma(2n+1)}}.
\end{equation}

Then, eq.~\ref{eq:radial_wf_overlap} becomes
\begin{equation}
\begin{aligned}
	& N_I N_F \int_0^\infty R^{n_I + n_f} \ee^{-\frac{\zeta_I + \zeta_F}{a_\mu}R} j_\lambda(Q R)\dd R \\
	=:& N_I N_F \int_0^\infty R^n \ee^{-\zeta R} j_\lambda(Q R)\dd R \\
	\label{eq:sto_ansatz}
\end{aligned}
\end{equation}
where we defined $n := n_I + n_F$ and $\zeta = (\zeta_I + \zeta_F)/a_\mu$. As derived in appendix~\ref{sec:int_j_lambda}, this integral can be evaluated analytically, yielding
\begin{multline}
	\langle j_\lambda(Q) \rangle = \frac{N_I N_F}{Q^{\lambda+2} (\zeta^2 + Q^2)^\frac{n-\lambda}{2}} \sum_{k=0}^{\frac{\lambda+1}{2}} \left( \frac{Q^2}{\zeta^2 + Q^2} \right)^k \\
	\left[
	A_{\lambda,n}^k Q \sin\left((2k+n-\lambda) \arctan\left(\frac{Q}{\zeta}\right)\right)
	+ \right.\\
	\left.
	B_{\lambda,n}^k \sqrt{\zeta^2 + Q^2} \cos\left((2k+n-\lambda - 1) \arctan\left(\frac{Q}{\zeta}\right)\right)
	\right],
	\label{eq:sto_result}
\end{multline}
with the constants $A_{\lambda,n}^k, B_{\lambda,n}^k$ as defined in appendix~\ref{sec:int_j_lambda}. This is the first main result.

\begin{table*}
\caption{Weighted radial wave function overlap in the STO model for monopole ($\lambda = 0$), dipole ($\lambda = 1$), and quadrupole ($\lambda = 2$) transitions. For the definitions of the constants $N_I, N_F, n, \zeta$ refer to the text.}
\label{tab:sto_result}
\begin{ruledtabular}
\begin{tabular}{ll}
	$\lambda=0$ & $\frac{N_I N_F \Gamma(n)}{Q (\zeta^2 + Q^2)^\frac{n}{2}} \sin\left(n \arctan\left(\frac{Q}{\zeta}\right)\right)$ \\
	$\lambda=1$ & $\frac{N_I N_F \Gamma(n-1)}{Q^2 (\zeta^2 + Q^2)^\frac{n}{2}} \left[ \zeta\sin\left(n \arctan\left(\frac{Q}{\zeta}\right)\right) - Q n \cos\left(n \arctan\left(\frac{Q}{\zeta}\right)\right) \right]$ \\
	$\lambda=2$ & \parbox{.95\textwidth}{\begin{multline}\textstyle\frac{N_I N_F \Gamma(n-2)}{Q^3 (\zeta^2 + Q^2)^\frac{n+1}{2}} \left[ \zeta (Q^2-3nQ^2-n^2Q^2+3\zeta^2) \sin\left(n \arctan\left(\frac{Q}{\zeta}\right)\right) - \right.\\\left. Q (Q^2+3n\zeta^2-n^2Q^2+3\zeta^2)\cos\left(n \arctan\left(\frac{Q}{\zeta}\right)\right) \right] \nonumber\end{multline}} \\
\end{tabular}
\end{ruledtabular}
\end{table*}

In table~\ref{tab:sto_result}, eq.~\ref{eq:sto_result} is evaluated for the most important transitions with $\lambda = 0, 1, 2$, corresponding to monopole, dipole, and quadrupole transitions, respectively.

It should be noted that for $Q \to \infty$, with
\begin{equation}
\begin{aligned}
	\sin\left[ \alpha \arctan\left( \frac{Q}{\zeta} \right)\right] &\approx \sin\left(\frac{\alpha\pi}{2}\right) - \frac{\alpha\zeta}{Q} \cos\left(\frac{\alpha\pi}{2}\right) \\
	\cos\left[ \alpha \arctan\left( \frac{Q}{\zeta} \right)\right] &\approx \cos\left(\frac{\alpha\pi}{2}\right) + \frac{\alpha\zeta}{Q} \sin\left(\frac{\alpha\pi}{2}\right) \\
\end{aligned}
\end{equation}
it is straight forward to derive that eq.~\ref{eq:sto_result} behaves as
\begin{multline}
	\frac{N_I N_F}{Q^{n + 1}} \left[
	\sin\left((n-\lambda) \frac{\pi}{2}\right)\sum_{k=0}^{\frac{\lambda+1}{2}} (-1)^k \left( A_{\lambda,n}^k + B_{\lambda,n}^k \right) -
	\right. \\ \left.
	\frac{\zeta}{Q} \cos\left((n-\lambda) \frac{\pi}{2}\right)\sum_{k=0}^{\frac{\lambda+1}{2}} (-1)^k \left( (2k+n-\lambda)A_{\lambda,n}^k + (2k+n-\lambda-1)B_{\lambda,n}^k \right)
	\right].
	\label{eq:sto_largeq}
\end{multline}
Note that the inclusion of the $\cos((n-\lambda)\pi/2)$ term is necessary in case $n$ is an integer and $n - \lambda$ is even. In that case, $\sin((n-\lambda)\pi/2) = 0$ and the asymptotic behavior is described by the second term only. Eq.~\ref{eq:sto_largeq} is very useful as it gives a simple approximation to the $Q$-dependence of $\langle j_\lambda(Q) \rangle$ for large arguments. This is important as it allows to easily determine the maximal $Q$ to be used in numerical simulations (or equivalently to estimate the systematic error introduced by considering only momentum transfers up to a certain maximum $Q$).

For $Q \to 0$ the behavior of eq.~\ref{eq:sto_result} is more generally more complicated\footnote{It is possible to do the expansion on a case--by--case basis, though, such as for those cases listed in tab.~\ref{tab:sto_result}. This yields the same results as derived from eq.~\ref{eq:sto_ansatz} by using the approximation for small $Q$, of course.}, though. On the one hand, $Q$ is nested deep inside several trigonometric functions, on the other hand many low-order terms cancel due to the unique form of the $A_{\lambda,n}^k, B_{\lambda,n}^k$. In addition, evaluating eq.~\ref{eq:sto_result} numerically is also dangerous (because of the division by small numbers). Hence, for this case, evaluating eq.~\ref{eq:sto_ansatz} for small $Q$ directly is favorable. Provided that $QR_\text{max} \ll \sqrt{\lambda+3/2}$ (where $R_\text{max}$ is the supremum of radii for which the product $\psi_I(R)\psi_F(R)$ is non-negligible), this becomes
\begin{equation}
\begin{aligned}
	&N_I N_F \int_0^\infty R^n \ee^{-\zeta R} j_\lambda(Q R)\dd R \\
	\approx & \frac{N_I N_F Q^\lambda}{(2\lambda + 1)!!} \int_0^\infty R^{n+\lambda} \ee^{-\zeta R} \dd R \\
	=& \frac{N_I N_F \Gamma(n+\lambda+1))}{\zeta^{n+\lambda+1}} \cdot \frac{Q^\lambda}{(2\lambda + 1)!!}
\end{aligned}
\label{eq:STO_small_q}
\end{equation}

This shows that the usually used approximation eq.~\ref{eq:jlambda_smallq} is perfectly recovered for small $Q$. Furthermore, eq.~\ref{eq:STO_small_q} is important for the actual implementation in simulation software packages. As was noted before, inserting $Q = 0$ into eq.~\ref{eq:sto_result} directly would produce a division-by-zero error. Hence, for $Q \approx 0$, eq.~\ref{eq:STO_small_q} should be used instead.

\section{Hydrogen-like orbitals}

Hydrogen-like orbitals are very similar to STO in so far as the same terms that appear for STOs appear in HLOs as well. There are four key differences, however: (a) in HLOs, the principal quantum number $n$ is an integer, (b) the radial part of the wave function depends on the angular momentum quantum number $l$, (c) HLOs have an additional factor represented by (generalized) Laguerre polynomials, and it can thus be ensured that (d) HLOs with same $l$, but different $n$ are orthogonal.

In general, the radial part of the HLOs can be written as
\begin{equation}
	N \ee^{-\frac{\zeta R}{n a_\mu}} \left( \frac{2\zeta R}{n a_\mu} \right)^l L_{n-l-1}^{2l+1}\left( \frac{2\zeta R}{n a_\mu} \right),
\end{equation}
where the $L_{n-l-1}^{2l+1}(x)$ are the generalized Laguerre polynomials, and $a_\mu = a_0 m_e / \mu$ with the electron mass $m_e$, the reduced mass $\mu$, and the Bohr radius $a_0$\footnote{For most practical applications, $\mu \approx m_e$ and hence $a_mu = a_0$ can be used}. Typically, $\zeta$ is set to be equal to the (unscreened) nuclear charge $Z$. This does not influence the further calculation here, however, and so we use $\zeta$ to indicate that screening may be included, and to preserve the analogy to the STO results. The normalization constant is given by
\begin{equation}
	N = \sqrt{\left( \frac{2\zeta}{n a_\mu} \right)^3 \frac{(n-l-1)!}{2n(n+l)!}}
\end{equation}

With this model, the weighted radial wave function overlap eq.~\ref{eq:radial_wf_overlap} becomes
\begin{equation}
\begin{aligned}
	\langle j_\lambda(Q) \rangle =& N_I N_F \int_0^\infty j_\lambda(Q R) \ee^{-\left( \frac{\zeta_I}{n_I a_\mu} + \frac{\zeta_F}{n_F a_\mu} \right) R} \\
	&R^2 \left(\frac{2\zeta_I R}{n_I a_\mu}\right)^{l_I}\left(\frac{2\zeta_F R}{n_F a_\mu}\right)^{l_F}  \\
	&L_{n_I-l_I-1}^{2l_I+1}\left( \frac{2\zeta_I R}{n_I a_\mu} \right) L_{n_F-l_F-1}^{2l_F+1}\left( \frac{2\zeta_F R}{n_F a_\mu} \right) \dd R,
\end{aligned}
\end{equation}
which can be rewritten as
\begin{equation}
	N_I N_F \sum_{b=l_I+l_F+2}^{n_I+n_F} p_b \int_0^\infty R^b \ee^{-\zeta R} j_\lambda(Q R) \dd R,
	\label{eq:hlo_ansatz}
\end{equation}
with $\zeta := \left( \frac{\zeta_I}{n_I a_\mu} + \frac{\zeta_F}{n_F a_\mu} \right)$ and the coefficients $p_b$ as defined in appendix \ref{sec:hlo_pb}.

The integral has exactly the same form as for STOs, with the solution (see appendix \ref{sec:int_j_lambda})
\begin{multline}
	\langle j_\lambda(Q) \rangle = \frac{N_I N_F}{Q^{\lambda+2}}
	\sum_{b=l_I+l_F+2}^{n_I+n_F} \frac{p_b}{(\zeta^2 + Q^2)^\frac{b-\lambda}{2}} \sum_{k=0}^{\frac{\lambda+1}{2}} \left( \frac{Q^2}{\zeta^2 + Q^2} \right)^k \\
	\left[
	A_{\lambda,b}^k Q \sin\left((2k+b-\lambda) \arctan\left(\frac{Q}{\zeta}\right)\right)
	+ \right.\\
	\left.
	B_{\lambda,b}^k \sqrt{\zeta^2 + Q^2} \cos\left((2k+b-\lambda - 1) \arctan\left(\frac{Q}{\zeta}\right)\right)
	\right].
	\label{eq:hlo_1}
\end{multline}
Since, contrary to the situation for STOs, $b$ is an integer here, this can be simplified further. With
\begin{equation}
\begin{aligned}
	\sin\left[ n \arctan\left( \frac{Q}{\zeta} \right)\right] &= \frac{\Im\left[ (\zeta+\ii Q)^n \right]}{(\zeta^2+Q^2)^{n/2}} \\
	\cos\left[ n \arctan\left( \frac{Q}{\zeta} \right)\right] &= \frac{\Re\left[ (\zeta+\ii Q)^n \right]}{(\zeta^2+Q^2)^{n/2}} \\ \end{aligned}
\end{equation}
one gets
\begin{multline}
	\langle j_\lambda(Q) \rangle = \frac{N_I N_F}{Q^{\lambda+2}}
	\sum_{b=l_I+l_F+2}^{n_I+n_F} \sum_{k=0}^{\frac{\lambda+1}{2}} \frac{p_b Q^{2k}}{(\zeta^2 + Q^2)^{2k+b-\lambda}} \\
	\left[
	A_{\lambda,b}^k Q \Im\left[ (\zeta+\ii Q)^{2k+b-\lambda} \right]
	+ B_{\lambda,b}^k (\zeta^2 + Q^2)
	\Re\left[ (\zeta+\ii Q)^{2k+b-\lambda-1} \right]
	\right].
	\label{eq:hlo_result}
\end{multline}
This is the second main result. It should be emphasized that eq.~\ref{eq:hlo_result} has the form of a the quotient of two polynomials,
\begin{equation}
	\frac{P(Q)}{Q^{\lambda+2}(\zeta^2 + Q^2)^{n_I + n_F + 1}}.
\end{equation}
As such, it is only marginally more complicated than the dipole approximation (which is simply a linear polynomial in $Q$), gives but an exact expression of the weighted radial wave function overlap in the framework of the HLO model.

\begin{table*}
\caption{Weighted radial wave function overlap in the HLO model for monopole ($\lambda = 0$), dipole ($\lambda = 1$), and quadrupole ($\lambda = 2$) transitions. For the definitions of the constants $N_I, N_F, n, \zeta$ refer to the text. The first column shows the spectroscopic notation, the second shows the orbitals involved.}
\label{tab:hlo_result}
\begin{ruledtabular}
\begin{tabular}{cccl}
	K & 1s~$\to$~2s & $\lambda = 0$ &
	$\frac{N_I N_F}{(Q^2 + \zeta^2)^2} \left[ 4\zeta + \frac{2(Q^2 - 3\zeta^2) \zeta_F}{a_\mu (Q^2 + \zeta^2)} \right]$ \\
	& 1s~$\to$~2p & $\lambda = 1$ &
	$\frac{N_I N_F Q}{(Q^2 + \zeta^2)^2}\left[ \frac{8 \zeta \zeta_F}{a_\mu (Q^2 + \zeta^2)} \right]$ \\
	L & 2s~$\to$~3s & $\lambda = 0$ &
	$\frac{N_I N_F}{(Q^2 + \zeta^2)^2} \left[ 12\zeta + \frac{2(Q^2-3\zeta^2)(3\zeta_I + 4\zeta_F)}{a_\mu(Q^2 + \zeta^2)} + \frac{16\zeta\zeta_F(\zeta^2-Q^2)(2\zeta_F+9\zeta_I)}{3a_\mu^2(Q^2 + \zeta^2)^2} - \frac{16\zeta_I\zeta_F^2(Q^4-10Q^2\zeta^2+5\zeta^4)}{3a_\mu^3(Q^2 + \zeta^2)^3} \right]$ \\
	& 2s~$\to$~3p & $\lambda = 1$ &
	$\frac{N_I N_F Q}{(Q^2 + \zeta^2)^2} \left[ \frac{128 \zeta \zeta_F}{3a_\mu(Q^2 + \zeta^2)} + \frac{64 \zeta_F (Q^2-5\zeta^2)(\zeta_F+3\zeta_I)}{9a_\mu^2(Q^2 + \zeta^2)^2} - \frac{64 \zeta\zeta_I\zeta_F^2 (3Q^2-5\zeta^2)}{3a_\mu^3(Q^2 + \zeta^2)^3} \right]$ \\
	& 2s~$\to$~3d & $\lambda = 2$ &
	$\frac{N_I N_F Q^2}{(Q^2 + \zeta^2)^2} \left[ \frac{128 \zeta \zeta_F^2}{3a_\mu^2(Q^2 + \zeta^2)^2} + \frac{64 \zeta_I \zeta_F^2 (Q^2-7\zeta^2)}{3a_\mu^3(Q^2 + \zeta^2)^3} \right]$ \\
	& 2p~$\to$~3s & $\lambda = 1$ &
	$\frac{N_I N_F Q}{(Q^2 + \zeta^2)^2} \left[ \frac{24 \zeta \zeta_I}{a_\mu(Q^2 + \zeta^2)} + \frac{16 \zeta_I \zeta_F (Q^2-5\zeta^2)}{a_\mu^2(Q^2 + \zeta^2)^2} - \frac{32 \zeta\zeta_I\zeta_F^2 (3Q^2-5\zeta^2)}{3a_\mu^3(Q^2 + \zeta^2)^3} \right]$ \\
	& 2p~$\to$~3p & $\lambda = 0$ &
	$\frac{N_I N_F}{(Q^2 + \zeta^2)^2} \left[ \frac{192 \zeta \zeta_I \zeta_F(\zeta^2 - Q^2)}{3a_\mu^2(Q^2 + \zeta^2)^2} - \frac{32 Q \zeta_I\zeta_F^2 (Q^4-10Q^2\zeta^2+5\zeta^4)}{3a_\mu^3(Q^2 + \zeta^2)^3} \right]$ \\
	& & $\lambda = 2$ &
	$\frac{N_I N_F Q^2}{(Q^2 + \zeta^2)^2} \left[ \frac{128 \zeta \zeta_I \zeta_F}{a_\mu^2(Q^2 + \zeta^2)^2} + \frac{64 \zeta_I\zeta_F^2 (Q^2-7\zeta^2)}{3a_\mu^3(Q^2 + \zeta^2)^3} \right]$ \\
	& 2p~$\to$~3d & $\lambda = 1$ &
	$\frac{N_I N_F Q}{(Q^2 + \zeta^2)^2} \left[ \frac{64 \zeta\zeta_I\zeta_F^2 (5\zeta^2-3Q^2)}{3a_\mu^3(Q^2 + \zeta^2)^3} \right]$ \\
\end{tabular}
\end{ruledtabular}
\end{table*}

In table~\ref{tab:hlo_result}, eq.~\ref{eq:hlo_result} is evaluated for the most important transitions with $\lambda = 0, 1, 2$, corresponding to monopole, dipole, and quadrupole transitions, respectively.

As before, one can study the behavior of eq.~\ref{eq:hlo_result} for the limit of $Q \to \infty$ and $Q \to 0$. For $Q \to \infty$, eq.~\ref{eq:hlo_1} reduces to
\begin{multline}
	N_I N_F \sum_{b=l_I+l_F+2}^{n_I+n_F} \frac{p_b}{Q^{b + 1}} \left[
	\sin\left( (b - \lambda) \frac{\pi}{2} \right) \sum_{k=0}^{\frac{\lambda+1}{2}} (-1)^k \left[ A_{\lambda,b}^k + B_{\lambda,b}^k \right] -
	\right. \\ \left.
	\frac{\zeta}{Q} \cos\left( (b - \lambda) \frac{\pi}{2} \right) \sum_{k=0}^{\frac{\lambda+1}{2}} (-1)^k \left[ (2k+b-\lambda)A_{\lambda,b}^k + (2k+b-\lambda-1)B_{\lambda,b}^k \right]
	\right].
	\label{eq:hlo_largeq}
\end{multline}
This exhibits exactly the same behavior as eq.~\ref{eq:sto_largeq}.

For $Q \to 0$, it is best to start from eq.~\ref{eq:hlo_ansatz}. Using eq.~\ref{eq:jlambda_smallq} (provided that $Q R_\text{max} \ll \sqrt{\lambda + 3/2}$ as before), one obtains
\begin{equation}
N_I N_F \frac{Q^\lambda}{(2\lambda + 1)!!} \sum_{b = l_I+l_F+2}^{n_I+n_F} \frac{p_b(b + \lambda)!}{\zeta^{b+\lambda+1}}
\label{eq:HLO_small_q}
\end{equation}
in accordance with predictions.

\section{Discussion}

Eqs.~\ref{eq:sto_result} and \ref{eq:hlo_result} are the main results of this work. Compared to the dipole approximation for small $Q$ (eqs.~\ref{eq:STO_small_q} and \ref{eq:HLO_small_q}), they offer a significant improvement. Unphysical cut-offs are no longer necessary to ensure they tend to zero properly for large $Q$. In addition, they are simple finite sums of polynomials and---in the case of STOs---elementary trigonometric functions. As such, they are easily implemented into existing simulation programs\cite{U_v110_i7_p831,Koch2002,Kirkland1998,U_v109_i4_p350} with very little effort and the increase in computation time is small.

\begin{figure}
\includegraphics{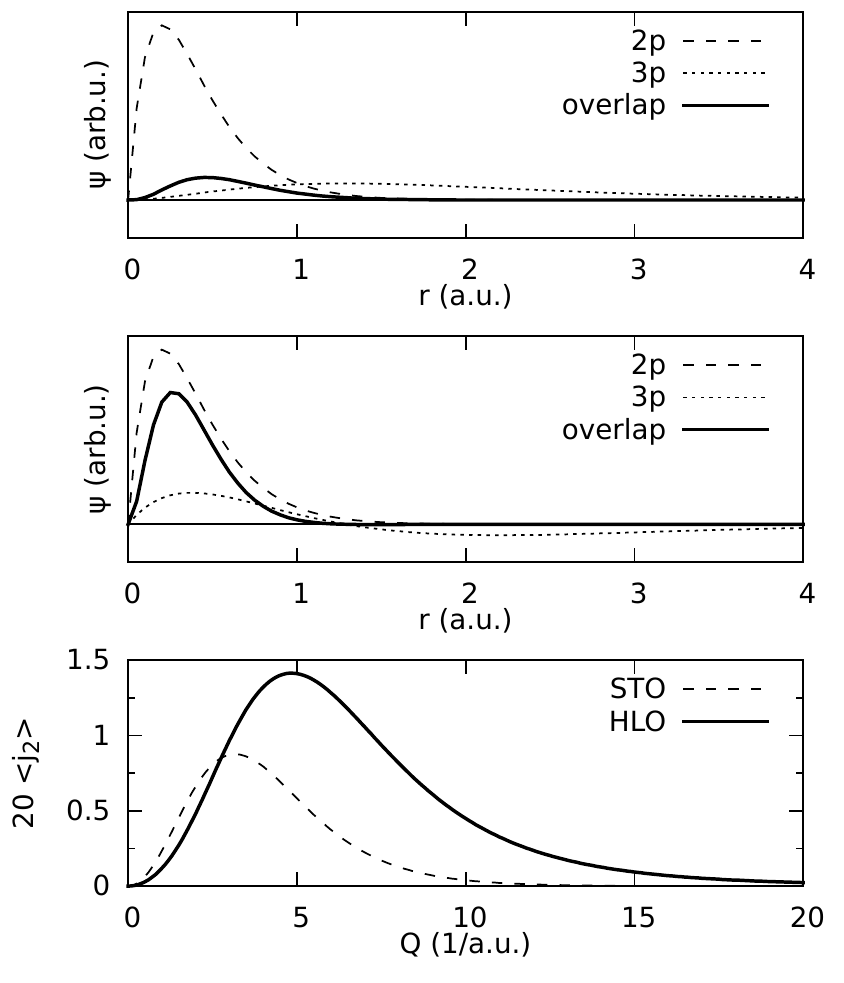}
\caption{Wavefunctions in the STO (top) and HLO (middle) models for 2p and 3p orbitals, as well as their overlap, together with the radial transition amplitude (bottom) for the quadrupole-allowed transition between those two states. Due to the missing node in the 3p wavefunction, the STO overlap is is shifted towards larger $r$ and underestimated in strength. Consequently, $\langle j_2(Q)\rangle$ is shifted towards smaller $Q$ and severely underestimated in the STO model.}
\label{fig:Si_2p_3p_STO_HLO}
\end{figure}

It must be emphasized that STOs are designed to have the same asymptotic behavior as HLOs, but no nodes. In the case of $l = n - 1$ (in which case the HLOs are also nodeless), they are identical. In all other cases, the missing nodes are problematic. This is particularly evident in the case of $l_i = l_f$ which can be allowed in monopole or quadrupole transitions. In those cases, STOs give completely wrong results because the node-less radial wave functions are not orthogonal as they should be, as can be seen from fig.~\ref{fig:Si_2p_3p_STO_HLO}.

\begin{figure}
\includegraphics{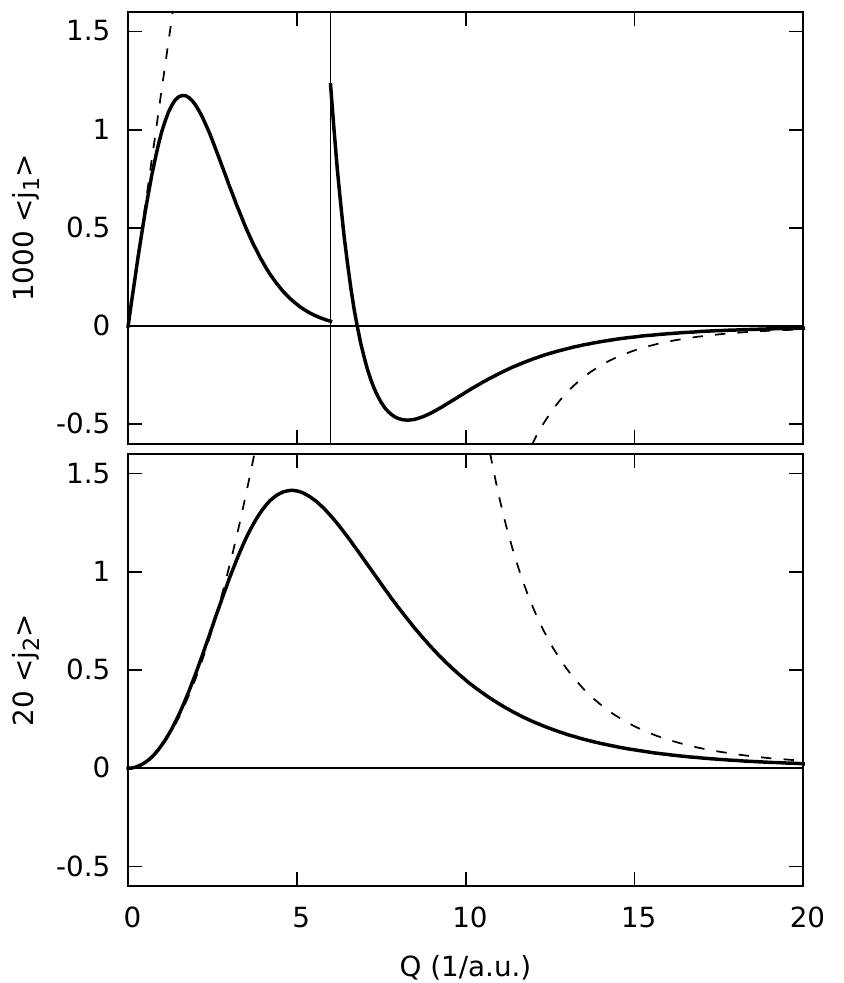}
\caption{Weighted radial wave function overlaps $\langle j_\lambda \rangle$ for the example of an isolated Si atom. The top panel shows a 2p~$\to$~3d dipole transition calculated using the STO model (for $Q \ge 6$, the curve is magnified by a factor of 50), the bottom panel shows a 2p~$\to$~3p quadrupole transition calculated using the HLO model.\footnote{The different orders of magnitude are caused by the different screening constants for the different final states. For the 3d state, the nuclear charge is almost completely screened, resulting in a redistribution of the orbital's probability density to larger radii, and hence to a reduction of the overlap with the initial state, compared to the transition to a 3p state. For the 2p~$\to$~3d transition, HLO and STO calculations are identical.} The screening constants for both cases were taken from ref.~\citenum{PR_v36_i1_p57}. The dashed lines show the asymptotic behavior.}
\label{fig:Si_2p_STO_HLO}
\end{figure}

In fig.~\ref{fig:Si_2p_STO_HLO}, examples of weighted radial wave function overlaps using both the STO and HLO models are shown in more detail. It is clearly evident that using the complete eqs.~\ref{eq:sto_result} and \ref{eq:hlo_result} gives highly superior results than the approximations for small or large $Q$ alone.

\begin{figure}
\includegraphics{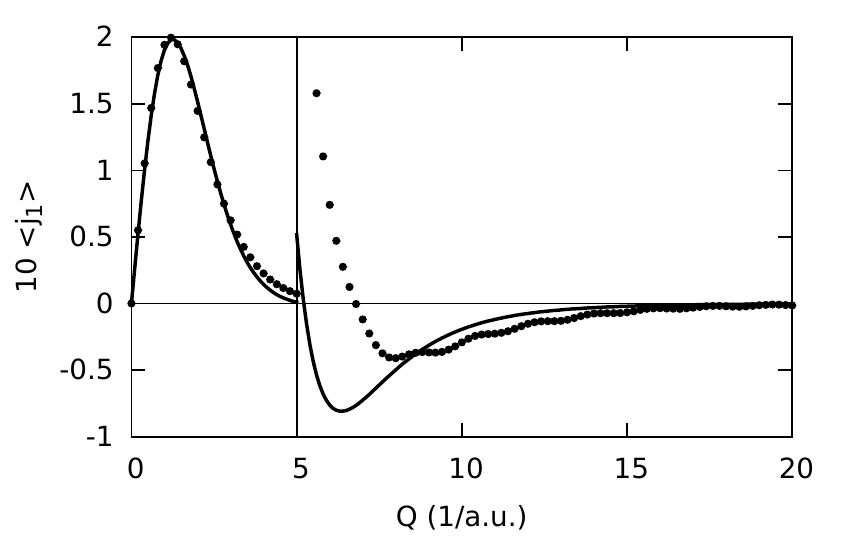}
\caption{Comparison of the 2p~$\to$~3d dipole-transition calculated for crystalline Si using the HLO model (solid line) and WIEN2k\cite{Wien2k} (dots). The screening constants for the HLO model were determined by fitting to the WIEN2k data. For $Q \ge 5$, the data is magnified by a factor of 50.}
\label{fig:wien2k}
\end{figure}

The formulas even give comparable results to much more sophisticated calculations using a full crystalline environment of the atom. In fig.~\ref{fig:wien2k}, a calculation using the HLO model is compared to WIEN2k\cite{Wien2k} calculations using the TELNES.3 program. The excellent agreement can primarily be attributed to the fact that the initial state---which has a high probability density in close proximity to the nucleus---can be viewed as a filter on the final state. Consequently, the weighted radial wave function overlap is dominated by the shapes of the orbitals close to the nucleus, and crystal effects like bonding play only a secondary role.

\section{Conclusion}

In this work, we have demonstrated that the transition probability in Coulomb scattering can be separated in an angular part and a radial part. The former is composed of well-known matrix elements of spherical harmonics. The latter can be expressed in simple independent-atom models such as the STO or the HLO model. Both yield simple algebraic expressions involving only finite sums of polynomials and (in the case of STOs) elementary trigonometric functions. Therefore, they are easy to implement into existing simulations programs without a large increase of the computation time, but with significant improvements in terms of accuracy, especially in the regime of medium momentum transfer.

Moreover, some of the weighted radial wave function overlaps have one or more zeros. Hence, they are suppressed at the corresponding momentum transfers, even though they are not forbidden by other selection rules. This can be exploited, e.g., for measuring faint signals from transitions that are normally hidden under a huge background from another transition with much higher transition probability.

Finally, the formulas presented here can be exploited in the future to experimentally determine properties of wave functions in atoms, like the screening effects of other electrons.

\begin{acknowledgements}
S.L. and P.S. acknowledge financial support by the Austrian Science Fund (FWF) under grant number I543-N20.
\end{acknowledgements}

\appendix
\section{Weighted integral of spherical Bessel functions}
\label{sec:int_j_lambda}

In this section, we calculate
\begin{equation}
\int_0^\infty R^n \ee^{-\zeta R} j_\lambda(Q R)\dd R.
\end{equation}

For this, we use the (finite!) expansion of the spherical Bessel functions in terms of trigonometric functions \cite{Abramowitz1965}
\begin{equation}
\begin{aligned}
	j_\lambda(x) =& \sum_{k = 0}^{\frac{\lambda+1}{2}} \frac{(-1)^{\lambda - k} 2^{\lambda-2k} (\lambda-k)!}{k!} \cdot \\
	&\left[
		\frac{\sin(x)}{x^{\lambda-2k+1}} \binom{-\frac{1}{2} - k}{\lambda-2k} - 2k \frac{\cos(x)}{x^{\lambda-2k+2}} \binom{-\frac{1}{2} - k}{\lambda-2k+1}
	\right] \\
	=:& \sum_{k = 0}^{\frac{\lambda+1}{2}} \left[ \tilde{A}_\lambda^k \frac{\sin(x)}{x^{\lambda-2k+1}} + \tilde{B}_\lambda^k \frac{\cos(x)}{x^{\lambda-2k+2}} \right]
\end{aligned}
\end{equation}
where the binomial coefficients have to be understood in a generalized sense, i.e.,
\begin{equation}
	\binom{n}{k} := \frac{\Gamma(n + 1)}{\Gamma(k + 1)\Gamma(n - k + 1)},
\end{equation}
with the gamma function $\Gamma(x)$. Since
\begin{equation}
	\Gamma\left(\frac{1}{2} - k\right) = \frac{(-4)^k k! \sqrt{\pi}}{(2k)!},
\end{equation}
the binomial coefficients can be simplified to conventional factorials, yielding
\begin{equation}
\begin{aligned}
	\tilde{A}_\lambda^k &= \frac{(-1)^k 2^{2k-\lambda} (2\lambda-2k)!}{(\lambda-2k)!(2k)!} \\
	\tilde{B}_\lambda^k &= \frac{(-1)^k 2^{2k-\lambda-1} k (2\lambda-2k+2)!}{(\lambda-k+1)(\lambda-2k+1)!(2k)!}.
\end{aligned}
\end{equation}
In particular, we have the special cases
\begin{equation}
\begin{aligned}
	\tilde{A}_\lambda^0 &= \frac{(2\lambda)!}{2^\lambda \lambda!} = (2\lambda - 1)!! &
	\tilde{B}_\lambda^0 &= 0 \\
	\tilde{A}_\lambda^{\frac{\lambda}{2}} &= (-1)^\frac{\lambda}{2} &
	\tilde{B}_\lambda^{\frac{\lambda}{2}} &= \frac{(-1)^\frac{\lambda}{2} \lambda (\lambda+1)}{2} \\
	\tilde{A}_\lambda^{\frac{\lambda+1}{2}} & = 0 &
	\tilde{B}_\lambda^{\frac{\lambda+1}{2}} & = (-1)^\frac{\lambda + 1}{2}
\end{aligned}
\end{equation}
and the general identities
\begin{equation}
\begin{aligned}
	\tilde{A}_\lambda^{k+1} &= -\frac{(\lambda-2k)(\lambda-2k-1)}{(\lambda-k)(2\lambda-2k-1)(k+1)(2k+1)} \tilde{A}_\lambda^k \\
	\tilde{B}_\lambda^{k+1} &= -\frac{(\lambda-2k+1)(\lambda-2k)}{k(2\lambda-2k+1)(\lambda-k)(2k+1)} \tilde{B}_\lambda^k \\
	\tilde{B}_\lambda^k &= \frac{k(2\lambda-2k+1)}{\lambda-2k+1} \tilde{A}_\lambda^k .
\end{aligned}
\end{equation}

With this, we can write\footnote{The integrals converge only if $2k + n - \lambda > -1$ (for the sine integral) and $2k + n - \lambda > 1$ (for the cosine integral). Since $n := n_I + n_F$, $k \ge 0$, and $\lambda < \max(n_I, n_F) = n_F$ without loss of generality, this translates to $n_I \ge -1$ for the sine integral, which is always fulfilled. Analogously, for the cosine integral we have $k \ge 1$ (since $\tilde{B}_\lambda^0 = 0$), and hence the resulting inequality is identical.}
\begin{equation}
\begin{aligned}
	& \int_0^\infty R^n \ee^{-\zeta R} j_\lambda(Q R)\dd R \\
	=& \sum_{k=0}^{\frac{\lambda+1}{2}} \left[
	\frac{\tilde{A}_\lambda^k}{Q^{n+1}} \int_0^\infty (QR)^{n+2k-\lambda-1} \ee^{-\frac{\zeta}{Q} (QR)} \sin(Q R)\dd (QR) +
	\right.\\
	&\left.\phantom{\sum_{k=0}^{\frac{\lambda+1}{2}} \Bigg[}
	\frac{\tilde{B}_\lambda^k}{Q^{n+1}} \int_0^\infty (QR)^{n+2k-\lambda-2} \ee^{-\frac{\zeta}{Q} (QR)} \cos(Q R)\dd (QR)
	\right] \\
	=& \frac{1}{Q^{\lambda+2} (\zeta^2 + Q^2)^\frac{n-\lambda}{2}} \sum_{k=0}^{\frac{\lambda+1}{2}} \left( \frac{Q^2}{\zeta^2 + Q^2} \right)^k \\
	&\left[
	A_{\lambda,n}^k Q \sin\left((2k+n-\lambda) \arctan\left(\frac{Q}{\zeta}\right)\right)
	+ \right.\\
	&\left.
	B_{\lambda,n}^k \sqrt{\zeta^2 + Q^2} \cos\left((2k+n-\lambda - 1) \arctan\left(\frac{Q}{\zeta}\right)\right)
	\right]
\end{aligned}
\end{equation}
with
\begin{equation}
\begin{aligned}
	A_{\lambda,n}^k &= \Gamma(2k+n-\lambda) \tilde{A}_\lambda^k \\
	B_{\lambda,n}^k &= \Gamma(2k+n-\lambda - 1) \tilde{B}_\lambda^k.
\end{aligned}
\end{equation}

\section{Polynomial coefficients of HLOs}
\label{sec:hlo_pb}
The polynomial part of the weighted radial wave function overlap using HLOs is given by
\begin{equation}
	R^2 \left( \frac{2\zeta_I R}{n_I a_\mu} \right)^{l_I} \left( \frac{2\zeta_F R}{n_F a_\mu} \right)^{l_F} L_{n_I-l_I-1}^{2l_I+1}\left( \frac{2\zeta_I R}{n_I a_\mu} \right) L_{n_F-l_F-1}^{2l_F+1}\left( \frac{2\zeta_F R}{n_F a_\mu} \right)
\end{equation}
By virtue of the series expansion of the associated Laguerre polynomials\cite{Abramowitz1965},
\begin{equation}
	L_n^\alpha(x) = \sum_{j = 0}^n (-1)^j \binom{n + \alpha}{n - j} \frac{x^j}{j!},
\end{equation}
and the Cauchy product of (finite) series, this can be written as
\begin{equation}
\begin{aligned}
	&\sum_{k=0}^{n_I+n_F-l_I-l_F-2} \sum_{j=0}^{k} R^{k+l_I+l_F+2} (-1)^k \binom{n_I+l_I}{n_I-l_I-1-j} \\
	&\binom{n_F+l_F}{n_F-l_F-1-k+j} \frac{\left( \frac{2\zeta_I}{n_I a_\mu} \right)^{l_I+j} \left( \frac{2\zeta_F}{n_F a_\mu} \right)^{l_F+k-j}}{j!(k-j)!} \\
	=&\sum_{b=l_I+l_F+2}^{n_I+n_F} R^b \cdot (-1)^{b+l_I+l_F} \left( \frac{2\zeta_I}{n_I a_\mu} \right)^{l_I} \left( \frac{2\zeta_F}{n_F a_\mu} \right)^{l_F} \\
	&\sum_{j=0}^{b-l_i-l_f-2} \binom{n_I+l_I}{n_I-l_I-1-j} \binom{n_F+l_F}{n_F+l_i+1-b+j} \\
	&\frac{\left( \frac{2\zeta_I}{n_I a_\mu} \right)^{j} \left( \frac{2\zeta_F}{n_F a_\mu} \right)^{b-l_i-l_f-2-j}}{j!(b-l_i-l_f-2-j)!},
\end{aligned}
\end{equation}
which is of the form
\begin{equation}
	\sum_{b=l_I+l_F+2}^{n_I+n_F} p_b R^b.
\end{equation}

%

\end{document}